\documentclass[sigconf, anonymous=False]{acmart}
\AtBeginDocument{%
  \providecommand\BibTeX{{%
    \normalfont B\kern-0.5em{\scshape i\kern-0.25em b}\kern-0.8em\TeX}}}

\setcopyright{acmcopyright}
\copyrightyear{2020}
\acmYear{2020}
\setcopyright{acmlicensed}\acmConference[BCB '20]{Proceedings of the 11th ACM International Conference on Bioinformatics, Computational Biology and Health Informatics}{September 21--24, 2020}{Virtual Event, USA}
\acmBooktitle{Proceedings of the 11th ACM International Conference on Bioinformatics, Computational Biology and Health Informatics (BCB '20), September 21--24, 2020, Virtual Event, USA}
\acmPrice{15.00}
\acmDOI{10.1145/3388440.3412477}
\acmISBN{978-1-4503-7964-9/20/09}

\usepackage{balance}
\usepackage{xspace}
\usepackage{enumitem} % an error message on enumitem (uncomment if in use)
\usepackage{xcolor,multirow, colortbl}
\usepackage{amsmath}
\DeclareMathOperator{\arccosh}{arcosh}
\PassOptionsToPackage{hyphens}{url}\usepackage{hyperref}

\newcommand{\modelname}{\texttt{Bio-JOIE}\xspace}
\newcommand{\covid}{{SARS-CoV-2}\xspace}

\newcommand{\todo}[1]{{\color{red}[Todo: \textbf{#1}]}}

\newcommand{\chelsea}[1]{{\color{blue}[CJ: \textbf{#1}]}}
\newcommand{\muhao}[1]{{\color{blue}[MC: \textbf{#1}]}}
\newcommand{\junheng}[1]{{\color{teal}[JH: {#1}]}}

\newcommand{\stitle}[1]{\vspace{0.3ex}\noindent{\bf #1}}

\usepackage{makecell}
\newcommand\newtoprule{\Xhline{.10em}}
\newcommand\newmidrule{\Xhline{.05em}}
\newcommand\newbottomrule{\Xhline{.10em}}
\newcommand{\tabincell}[2]{\begin{tabular}{@{}#1@{}}#2\end{tabular}}
\colorlet{shadecolor}{gray!15}

\begin{document}

\title[Joint Representation Learning of Biological Knowledge Bases]{\modelname: Joint Representation Learning of Biological Knowledge Bases}


\author{Junheng Hao$^1$, Chelsea J.-T Ju$^1$, Muhao Chen$^2$, Yizhou Sun$^1$, Carlo Zaniolo$^1$, Wei Wang\,$^1$}
\affiliation{
$^{\text{\sf 1}}$Department of Computer Science, University of California Los Angeles, Los Angeles, CA, 90095, USA\\
$^{\text{\sf 2}}$Department of
Computer and Information Science, University of Pennsylvania, Philadelphia, PA 19104, USA
}
\email{[jhao,chelseaju,yzsun,zaniolo,weiwang]@cs.ucla.edu,muhao@seas.upenn.edu}
\renewcommand{\shortauthors}{Hao, et al.}

\begin{abstract}
% \junheng{Current abstract match the introduction but not the title. Alternative way: The revised abstract does not start with \covid story but mentions at the application.} \red{Abstract long.}
%\todo{recheck according to intro. The abstract should not exceed 250 words!} %The outbreak of the COVID-19 respiratory disease has led to a worldwide pandemic with high mortality. 
%Intense efforts by virologists have focused on understanding its causative agent (named \covid)
%including the virus-human protein-protein interactions (PPIs) and gene ontology annotations for viral proteins. 
%The knowledge in the domains of proteins and gene ontology are fundamental to many downstream applications \junheng{such as PPI classification, however, one single domain is not enough.}

The widespread of Coronavirus has led to a worldwide pandemic with a high mortality rate. 
%Intense efforts have been devoted by scientists to understand the causative agent and to elucidate its molecular impact on human. 
Currently, the knowledge accumulated from different studies about this virus is very limited. 
Leveraging a wide-range of biological knowledge, such as gene ontology and protein-protein interaction (PPI) networks from other closely related species presents a vital approach to infer the molecular impact of a new species. 
%\covid related protein interaction predictions are essential to disclose the mechanism of the human infection process and beneficial to downstream applications such as identifying drug targets and repurposing.
%In fact, most other species have similar multi-domain 
%\YS{what does domain mean here? Is it a proper word?} 
%biological knowledge bases that represent different aspects of their biological mechanism, and missing knowledge can be transferred from other KBs and  
%\WW{what do you mean by "domain"?}
%However, existing methods  \WW{methods for what tasks?} focus on feature extraction and learning from one specific domain and fall short of utilizing complementary knowledge from others. \WW{I do not follow what you are trying to convey here} 
In this paper, we propose the transferred multi-relational embedding model \modelname to capture the knowledge of gene ontology and PPI networks, which demonstrates superb capability in modeling the \covid-human protein interactions.
\modelname jointly trains two model components. 
The \emph{knowledge model} encodes the relational facts from the protein and GO domains into separated embedding spaces,
using a hierarchy-aware encoding technique employed for the GO terms.
On top of that, the \emph{transfer model} learns a non-linear transformation to transfer the knowledge of PPIs and gene ontology annotations across their embedding spaces.
By leveraging only structured knowledge, \modelname significantly outperforms existing state-of-the-art methods in PPI type prediction on multiple species. %without using  information such as protein sequences and descriptions of GO terms.
%Furthermore, the learned protein representations from \modelname can also be used for unsupervised clustering to detect protein families. 
Furthermore, we also demonstrate the potential of leveraging the learned representations on clustering proteins with enzymatic function into enzyme commission families.
%\WW{what do you mean by identifying protein families?}\junheng{This is related to one task doing clustering on embeddings, following Onto2Vec paper.
Finally, we show that \modelname can accurately identify PPIs between the \covid proteins and human proteins, providing valuable insights for advancing research on this new disease. 
% Source code and preprocessed datasets are available at: \href{https://www.haojunheng.com/project/goterm}{https://www.haojunheng.com/project/goterm}.
\end{abstract}




\maketitle

\section{Introduction}\label{sec:introduction}
% \junheng{Para. \#1: Introduction to COVID-19 and virus-human protein importance}
% \todo{Global change: (1) ``complementary'' as ``augmented''? (2) Expand ``GO'' to gene ontology.} \todo{necessities to learn multi-domain knowledge}
The outbreak of COVID-19 (Coronavirus Disease-2019) %, a severe respiratory disease, 
has infected over millions of people and caused high death tolls since the end of 2019, as worldwide social and economic disruption.
%\footnote{Data source: \href{https://coronavirus.jhu.edu/map.html}{https://coronavirus.jhu.edu/map.html}, updated as 6pm (ET), 04/29/2020. }
%\YS{do you want to change to the latest number?} \WW{I would suggest that we rephrase this sentence to remove explicit mentioning of the actual number, because any accurate numbers now will soon become obsolete before the audience reads this paper.}
Tremendous efforts have been made to discover the infection mechanism of the causative agent, named \covid. 
%to design a possible cure and immunity. 
%Biologists and virologists worldwide are working around the clock to %contribute relevant scientific knowledge
%assemble resources of scientific knowledge to help combat this severe respiratory disease.
One important and urgent task is to understand the mechanism in which viral proteins interact with human proteins. 
%\todo{not major task in our paper?}
%The collection of such resources includes finding existing or suspected interactions between viral proteins and human proteins, 
The new findings will  enrich the annotation of viral genomes \citep{gordon2020sars} in biomedical knowledge bases (KBs). 
%as shown in Figure \ref{fig:covid-ppi-example} and \ref{fig:covid-go-example}.
%Such newly curated biological knowledge bases on protein and gene ontologies, together with established ones,
Constructing and populating such biomedical KBs can significantly improve our understanding of the processes by which \covid affects different cells in human body and will serve as the foundation for many important downstream applications such as vaccine development \cite{kamminga2019risk}, drug repurposing~\citep{zhou2020network,gordon2020sars} and drug side effect detection \cite{zitnik2018modeling}. 
%For example, if new interactions are observed between viral proteins and human proteins (such as the fact that \covid enter the human body through ACE2), scientists may likely use the fact in the KB to develop the cure \cite{zheng2020covid}.
%For example, recent work has demonstrated the use of AI systems to identify targets and potential therapeutics by querying the knowledge base, therefore facilitating rapid drug development \cite{stebbing2020covid}.

\begin{figure}[!ht]
    \centering
    \includegraphics[width=0.9\columnwidth]{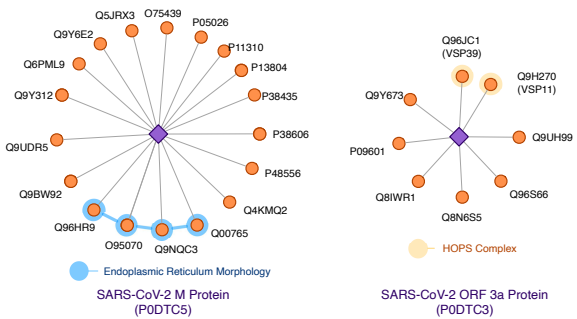}
    \caption{Two examples of \covid-human protein interactions: M protein (left) and ORF3a protein (right). {The purple diamonds refers to the viral proteins and the orange circles refer to the high-confidence human protein target. Proteins highlighted in blue are involved in certain biological processes, and proteins highlighted in yellow are arranged in a protein complex.}}
    \label{fig:covid-ppi-example}
\end{figure}

\begin{figure}[!ht]
    \centering
    \includegraphics[width=0.9\columnwidth]{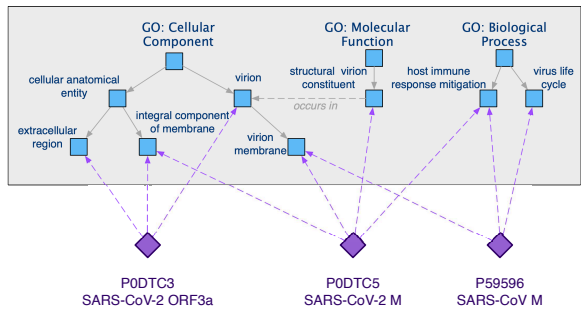}
    \caption{Examples of gene ontology annotation enrichment on three representative SARS-CoV or \covid proteins, which possess multiple properties across three biological aspects: biological processes, cellular components and molecular functions.}
    \label{fig:covid-go-example}
\end{figure}

% \junheng{Para. \#2: Existing knowledge bases that can be utilized to predict PPI}
% \chelsea{Note: 1. one or two sentences introduce biological knowledge bases, and their importance. \\
% 2. Give a few examples of these knowledge bases, and try to group them into different categories. Also briefly discuss what we can use these knowledge bases for. \\
% example - proteomic domain: STRING provides the protein-protein interaction information, Pfam provides information on protein domain and protein family. \\
% example - genomic domain: gene ontology (GO) knowledgebase provides a comprehensive source of information on the functions of genes; ClinVar aggregates information about genomic variation and its relationship to human health \\ 
% example - metabolic domain: BioCyc and KEGG provide the metabolic pathway information. BRENDA - a comprehensive enzyme information system \\
% example - expression domain: Expression Atlas, an open public repository of gene expression pattern data under different biological conditions. CaGE: Cardiac gene expression knowledgebase. \\
% example - drug domain: drug bank combines detailed drug data with comprehensive drug target information
% 3. Discuss the importance of cross-referencing this knowledge base (interrelated across different perspectives).
% }

In general, biological KBs, often stored as knowledge graphs (KGs), consist   of various biological entities, their properties and relations. These KBs can be categorized in different domains, such as gene annotation, functional proteomic analysis, and transcriptomic profiling. Specifically, gene ontology (GO) \citep{gene2018gene,huntley2015goa} is the most widely used resource for gene function annotation; STRING \cite{szklarczyk2016string}, PDB \cite{berman2007worldwide} and neXtProt \cite{lane2012nextprot} collect the knowledge accumulated from functional proteomic analysis; Expression Atlas~\cite{papatheodorou2020expression} is a database facilitating the retrieval and analysis of gene expression studies.
%In general, biological KBs are defined in various categories, %such as MESH \todo{ref}, DrugBank~\citep{wishart2018drugbank}, RCSB PDB~\citep{makrodimitris2019sparsity} and AmiGO ~\citep{carbon2009amigo}, 
%including proteomic databanks \citep{szklarczyk2016string,berman2007worldwide,lane2012nextprot},
%gene ontologies \citep{gene2018gene,huntley2015goa} and microarray knowledge bases \citep{kolesnikov2015arrayexpress,edgar2002gene,demeter2007stanford},
%store large knowledge graphs (KGs) for  
%various biological entities, and record their properties, structures and interaction information.
%Therefore, they provide the essential sources of knowledge for in silico research of biology and healthcare, and help tasks such as {therapeutic target identification~\citep{} and polypharmacy side effect detection} \citep{zitnik2018modeling} % \muhao{drugs do not belong to biological research}. \todo{add other representative applications, if necessary}.
%Many KBs in biology are (semi)structured, multi-relational, externally connected and interrelated \junheng{not sure} with other KBs. 
While those KBs provide the essential sources of knowledge for \textit{in silico} research in the corresponding domains,
such domain-specific knowledge is often sparse and costly to apprehend~\cite{makrodimitris2019sparsity,thomas2012use}. 
%\chelsea{Elaborate the issue of sparsity here.} %\junheng{Added yeast.}
%For example, PPI networks are far from well-populated \cite{huang2018completing,makrodimitris2019sparsity}. 
%Even for yeast, one of the best characterized model species with densest PPI networks, have only roughly 0.6\% of all possible pairs of proteins that are known to interact~\cite{makrodimitris2019sparsity}.
For example, PPI networks can be far from complete given the information supported by experimental results or suggested by computational inference~\cite{huang2018completing,makrodimitris2019sparsity}. \citet{makrodimitris2019sparsity} indicate that the numbers of PPIs in BIOGRID~\cite{oughtred2019biogrid} for non-model organisms are far less than expected, specifically, there are only 107 interactions for tomato (\textit{Solanum lycopersicum}) and 80 interactions for pig (\textit{Sus scrofa}). Evidently, relying on the KG from a single domain presents the risk of learning from limited and scarce information.

%On the other hand,
The stored knowledge is often interrelated across different perspectives. Hence, the missing knowledge in certain KBs can be transferred from other KBs, and thus provide a more comprehensive representation of the biological entities. 
%Protein-protein interactions (PPIs), protein annotations on biological processes, together with gene ontology itself can be integrated in knowledge bases and multiple types entities (proteins, GO terms, etc) with multiple relations are often well interrelated to form cross-domain knowledge and provide a comprehensive understanding of all biological entities, particularly those related to \covid. 
Taking the protein-protein interaction (PPI) examples of the new \covid proteins as illustrated in Figure~\ref{fig:covid-ppi-example}, \covid M protein interacts with a list of human proteins, and five of them are involved in the endoplasmic reticulum (ER) morphology process as suggested by the gene ontology annotation (GO:0005783).
% \junheng{GO\_0005783:endoplasmic reticulum. The conclusion is from the \covid nature paper.}. 
Similarly, the \covid ORF3a also interacts with a list of human proteins. Among these proteins, VSP39 and VSP11 are the core subunits of HOPS complex, presenting a binding action as suggested by the STRING database. 
%\junheng{\covid ORF3a interacts proteins which are HOPS complex as shown in Figure 1.  The idea is that, \covid interacts with proteins that are possibly related to a biological process or protein complex.}.
While aligning the gene ontology annotations of the \covid M protein as demonstrated in Figure~\ref{fig:covid-go-example}, the SARS-CoV M protein presents a similar set of gene ontology annotations, such as ``host immune mitigation'' and ``virion membrane'', suggesting that the side knowledge of gene ontology annotations can facilitate the inference of interactions for related proteins.
%Take the PPI and GO annotations of the aforementioned new \covid as one example. 
%In Figure \ref{fig:covid-ppi-example}, \covid M protein has a group of transactions with human proteins related to the ER morphology process and \covid ORF3a interacts with proteins from the same protein complex. 
%In addition, the M proteins from both \covid and SARS-CoV with similar GO annotations (such as ``host immune mitigation'') suggest a high similarity from previously known viruses, as shown in Figure \ref{fig:covid-go-example}.
More generally, the sparse domain information can always benefit from the supplementary knowledge from other relevant domains, therefore calling upon a plausible method to support the fusion and transfer of knowledge across multiple biological domains.
%Such observations will no doubt help discover more target proteins on \covid from the cross-domain knowledge. 

% \junheng{Para. \#3: Benefit of representation learning and challenges.}
%\chelsea{This section seems to be dangling here now...}
Regardless of the importance and advantages of knowledge fusion across different domains \citep{bleiholder2009data,bryl2014learning},
fewer efforts have been devoted to incorporating knowledge from different domains for a specific task in computational biology studies. 
Onto2vec~\cite{smaili2018onto2vec} presents one state-of-the-art learning approach that successfully bridges gene ontology annotations with the protein representation. However, the known PPI information is neglected and not encoded in the obtained protein embeddings.

To combine multiple domain-specific biological knowledge, and facilitate knowledge transfer across different domains, we purpose \modelname, a \underline{J}o\underline{I}nt \underline{E}mbedding learning framework for multiple domains of \underline{Bio}logical KBs.
In \modelname, %given multiple domains and associations among different domains in the KBs,
two model components are jointly learned, i.e.,
a knowledge model characterizes different domain-specific KGs in separate low-dimensional embedding spaces, and a transfer model captures the cross-domain knowledge association. 
More specifically, the knowledge model encodes the relational facts of entities in each view into the corresponding embedding space separately, with a hierarchy-aware technique designated for the hierarchically-layered domains. 
Besides, the transfer model seeks to transfer the knowledge between pairs of domains by employing a weighted non-linear transformation across their embedding spaces. 
% \junheng{Para. \#5: Contributions.}  Our contributions are 4-fold. 
%\todo{1. model. 2. enhancement. 3. multi-species and inductive adaptation. 4. unsupervised pattern visualization.}
In evaluation, we apply the \modelname on several PPI networks with Gene Ontology annotations and the entire gene ontology and evaluate by PPI predictions.
We compare \modelname with that of the state-of-the-art representation learning approaches on multiple species, including \covid-Human PPIs, with different model settings. Our best \modelname outperforms alternative approaches by 7.4\% in PPI prediction.

Our contributions are 4-fold. 
First, we construct a general framework for learning representations across different domain-specific KBs, including the dynamically changing \covid KB. 
Second, we emphasize and demonstrate that cross-domain representation learning by the proposed \modelname can improve the inference in one domain by leveraging the complementary knowledge from another domain. Extensive experiments on different species confirm the effectiveness of cross-domain representation learning.
Third, \modelname also demonstrates cross-species transferability to improve PPI predictions among multiple species by knowledge population from  gene ontology.
%It is also adaptive for training from the knowledge emerging \covid based on the connection to existing domains.
%Fourth, learning embeddings from \modelname are useful for unsupervised tasks such as clustering. We show that the protein representations from \modelname trained on PPI networks and gene ontology can also be used to help better identify protein families. 
%\todo{rephrase the contribution bullet points based on experiments}.
Fourth, the protein representations learned from \modelname can be leveraged for different tasks. Specifically, we show that the protein embeddings trained on PPI network and gene ontology present the potential to better group enzymes into different enzyme commission families. Tremendous efforts have been made to discover the infection mechanism of the causative agent, named \covid.

\section{Materials and Method} \label{sec:method}

In this section, we present the proposed method to support representation learning and cross-domain knowledge transfer on biological KBs. 
Without loss of generality and aligned with the evaluation of the proposed \modelname, we refer two domain-specific KGs in the following section to PPI networks and the gene ontology graph.
We begin with the formalized descriptions of the materials and tasks. 
\subsection{Preliminary} \label{subsec:prelimiary}

\stitle{Materials}. 
A typical biological KB can be viewed as relational data that are presented as an edge-labeled directed graph $\mathcal{G}$,
which is formed with a set of entities (e.g. proteins) $\mathcal{E}$ and a set of relations (e.g. interaction types) $\mathcal{R}$.
A triple $(s, r, t) \in \mathcal{G}$ represents a $r\in\mathcal{R}$ typed relation between the source and target entities $s, t \in \mathcal{E}$,
As stated, we continue with the modeling on KGs of two domains, PPI and gene ontology. 
%\junheng{add examples from protein and GO}
For example, in the PPI network, a triple \texttt{(FBgn0011606, binding, FBgn0260855)} simply states the fact that two proteins (from fly) have binding interaction; and in gene ontology, a triple \texttt{(GO:0008152, is a, GO:0008150)} similarly represents that GO:0008152 (a unique identifier of ``metabolic process'') is one subclass of GO:0008150 (a unique identifier of ``biological process'').
Our model seeks to capture the protein information in the triples $(s_p, r_p, t_p)$  of PPI graph $ \mathcal{G}_p$ in a $k_p$-dimensional embedding space,
where we use boldfaced notations such as $\mathbf{s}_p, \mathbf{r}_p, \mathbf{t}_p \in \mathbb{R}^{k_p}$ to denote the embedding representation.
Similarly, gene ontology is another graph $\mathcal{G}_o$ formed with a set of GO terms $\mathcal{E}_o$ and a set of semantic relations $\mathcal{R}_o$.
The triple $(s_o, r_o, t_o) \in \mathcal{G}_o$ identifies a semantic relation of GO terms, while we also observe hierarchical substructures formed by ``subclass'' or ``is\_a'' relation as the aforementioned example.
The gene ontology is embedded in another space $\mathbb{R}^{k_o}$, such that $k_p$ and $k_o$ may not be equivalent.
We use $(o,p) \in \mathcal{A}$ to denote a \emph{GO term annotation} where a GO term $o\in \mathcal{E}_o$ describes a protein $p\in \mathcal{E}_p$ of its corresponding functionality, and $\mathcal{A}$ denotes the set of such associations.
As introduced in Section \ref{sec:introduction}, we consider \covid-Human interaction as a similar (but significantly smaller) KBs with the same structures as $\mathcal{G}_p$, which serves as an extension of human PPI networks. 

\stitle{Tasks.} To validate the learned embedding of biological entities (proteins and GO terms in this context), we address the following two tasks.
(i) \emph{PPI type prediction} aims at predicting the interaction type between two interacting proteins, including \emph{\covid related PPIs};
(ii) \emph{Protein clustering and family identification} aims at clustering the existing proteins and helps identify the clusters based on Enzyme Commission (EC) numbers.

\begin{figure}[!ht]
    \centering
    \includegraphics[width=0.95\columnwidth]{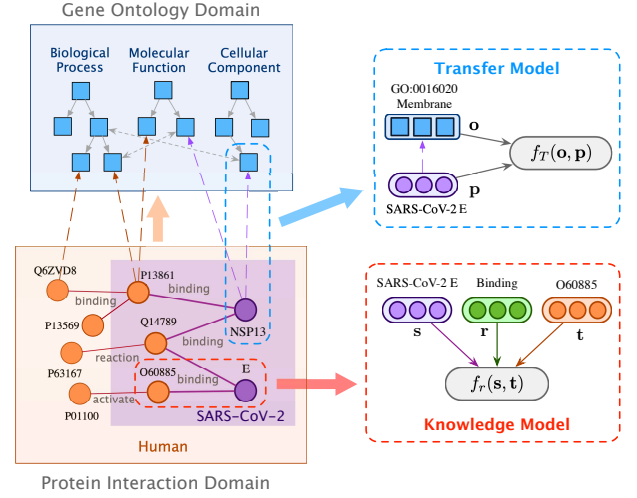}
    \caption{Model architecture of \modelname. The Knowledge Model seeks to encode relational facts in each domain respectively (such as proteins and gene ontology). Meanwhile, the Transfer Model learns to connect both domains and enable knowledge transfer across protein and gene ontology.}
    \label{fig:intro-goterm}
\end{figure}

\stitle{Methods}.
The model architecture of \modelname is shown in Figure \ref{fig:intro-goterm}. The proposed \modelname jointly learns two types of model components to connect the two views of structured knowledge.
{Knowledge models} are responsible for representing the relational knowledge of PPI and that of GO term into two separate embedding spaces $\mathbb{R}^{k_p}$ and $\mathbb{R}^{k_o}$ by using KG embedding and hierarchy-aware regularization.
On top of that, a {transfer model} learns a transformation to connect between the representations of GO term relation facts and PPI based on partially provided GO term assignments.
In particular, we investigate weighted \emph{transfer techniques} to better capture the knowledge transfer, for which the weights reflect the specificity of the assigned GO term to a protein. 
The following of this section describes the model components and the learning objective of \modelname in detail.

% \todo{Optional: model architecture.}
% \begin{figure}[!ht]
%     \centering
%     \includegraphics[width=0.9\columnwidth]{figures/model-archi.eps}
%     \caption{ Model architecture of \modelname.}
%     \label{fig:covid-ppi-percentage}
% \end{figure}

\subsection{Knowledge Model} \label{subsec:knowledge_model}

The knowledge models seek to characterize the semantic relations of GO terms and PPI information into separate embedding spaces.
In each embedding space,
the inference of relations or interactions is modeled as specific algebraic vector operations.
As mentioned, the two views of gene ontology and PPI are embedded to separate embedding spaces.

To capture a triple $(s,r,t)$ from either of the two domains,
a cost function $f_r(s,t)$ is provided to measure its plausibility.
A lower score indicates a more plausible triple.
We can adopt multiple vector operations in the defined embedding space with three representative examples defined as follows, i.e. translations (TransE~\citep{bordes2013translating}), Hadamard product \citep{yang2014embedding} and circular correlation (HolE~\citep{nickel2016holographic}). The cost functions are given as follows, where the symbol $\circ$ denotes Hadamard product, and
$\star:\mathbb{R}^{d}\times\mathbb{R}^{d}\rightarrow\mathbb{R}^{d}$ denotes circular correlation defined as $[\mathbf{a} \star \mathbf{b}]_k = \sum^{d}_{i=0} a_{i}b_{(k+i)\mod d}$.

\begin{equation*}
    \begin{aligned}
        f_r^{\mathrm{Trans}}(\mathbf{s},\mathbf{t}) & =  ||\mathbf{s} + \mathbf{r} - \mathbf{t}||_2\\
        f_r^{\mathrm{Mult}}(\mathbf{s},\mathbf{t}) & = - (\mathbf{s}\circ\mathbf{t})\cdot\mathbf{r}\\
        f_r^{\mathrm{HolE}}(\mathbf{s},\mathbf{t})  & = - (\mathbf{s} \star \mathbf{t})\cdot\mathbf{r}
    \label{equ:triple_loss}
    \end{aligned}
\end{equation*}
Since most of the relations in PPI networks are symmetric (such as binding and catalysis), we apply the Hadamard product based function.
The learning objective of a knowledge model on a graph $G$ is to minimize the following margin ranking loss,
\begin{equation*}
    \mathcal{L}^\mathcal{G}_K=\frac{1}{|\mathcal{G}|}\sum_{(s,r,t)\in \mathcal{G}} \max\left \{ f_r(s,t) + \gamma^\mathcal{G} - f_r(s',t'),\; 0 \right \}
\end{equation*}
where $\gamma^\mathcal{G}$ is a positive margin, and a negative sample $(s',r,t')\notin \mathcal{G}$ is created by randomly substituting either $s$ or $t$ using Bernoulli negative sampling \citep{wang2014knowledge}. 
With regard to the two domains of relational knowledge (proteins and gene ontology) $\mathcal{G}_p$ and $\mathcal{G}_o$, we denote the learning objective losses as $\mathcal{L}^{\mathcal{G}_p}_K$ and $\mathcal{L}^{\mathcal{G}_o}_K$.

\stitle{Hierarchy-aware Encoding Regularization}\label{sec:hier_reg}
As mentioned in Section \ref{subsec:prelimiary}, it is observed that some ontological knowledge can form hierarchies \cite{chen2018on2vec}, which is typically constituted by a relation with the implicit hierarchical property, such as ``subclass\_of'', as substructures. In gene ontology, more than 50\% of the triples have such relations.
To better characterize such hierarchies, we model such substructures differently from the aforementioned DistMult and many others by adding hierarchy regularization. More specifically, given entity pairs
$(e_l, e_h)\in S$ where $e_l$ is a subclass of $e_h$, we model such hierarchies by minimizing the distance 
between coarser concepts and associated finer concepts in embedding space. Hence, the loss is simply defined as
\begin{equation*}
    \mathcal{L}_{(\text{HA})} = \frac{1}{|S|}\sum_{(e_l, e_h)\in S} \left[ ||e_l - e_h||_2-\gamma_{\text{HA}} \right]_+
\end{equation*}
where $[x]_+=\max\{x,0\}$ and $\gamma_{\text{HA}}$ is also a positive margin parameter. This penalizes the case where the embedding of $e_l$ falls out the $\gamma_{\text{HA}}$-radius neighborhood centered at the embedding of $e_h$.

% \stitle{Graph neural network based Encoding} \todo{specific form of GNN to be added or decided.}

\paragraph{Relation Inference}
Given the learned embeddings and a pair of query proteins ($(p_1, p_2)$), we can predict the most plausible interaction type $r$ by selecting the optimal $f_r(p_1, p_2)$ score. We can also provide predictions for possible protein targets given the query of the subject protein and specific interaction type $(p, r, ?t)$ by populating the selection proteins with top score $f_r(p, t)$ from the knowledge model.
Details about each task are curated in Section \ref{subsec:ppi} and \ref{subsec:cov_ppi}.

\subsection{Transfer Model} \label{subsec:transfer_model}

The transfer model learns to connect between the above two relational embedding spaces via a non-linear transformation.
The transformation is induced based on the GO term assignments,
towards the goal to collocate the associated GO terms and proteins in an embedding space after transformation.
Hence, the affinity of embedding structures of gene ontology and PPIs can be captured.
This allows the relational knowledge to transfer across and complement the learning and inference on both domains.

Given each GO term assignment $(o,p)\in \mathcal{A}$,
following function $f_T(o,p)$ measures the plausibility of the transformation that is favored to be minimized.
\begin{equation*}
    f_T(o,p)=\left \| \sigma \left ( \mathbf{M}_T\cdot\mathbf{p} + \mathbf{b}_T \right ) - \mathbf{o} \right \|_2
\end{equation*}
$\mathbf{M}_T\in \mathbb{R}^{k_o \times k_p}$ thereof is a weight matrix and $\mathbf{b}_T\in \mathbb{R}^{k_p}$ is a bias vector.
$\sigma$ is either the identify function, or a non-linear function as $\mathrm{tanh}$,
the latter thereof aim at smoothing the transformation with additional non-linearity.

\subsubsection{Basic Transfer Model}

The basic strategy to learn the transfer model is to treat each GO term assignment evenly, and thereby minimizing the following learning objective loss.
\begin{equation*}
    \mathcal{L}_{T_1}=\frac{1}{|\mathcal{A}|}\sum_{(o,p)\in \mathcal{A}}\max \left \{ f_{T_1}\left ( o,p \right ) + \gamma^\mathcal{A} - f_{T_1}\left ( o',p' \right ) , \; 0 \right \}
\end{equation*}
$(p',o')\notin \mathcal{A}$ thereof is a negative sample by randomly substituting $p'$, and $\gamma^\mathcal{A}$ is a positive margin.

\subsubsection{Weighed Transfer Model}
Since some ontological knowledge, such as gene ontology, may form hierarchical structures, where GO terms in lower levels typically describe more specified gene functionality.
During the characterization of associations between GO terms and proteins, in contract to general GO terms,
more specified GO terms necessarily carry more precise descriptions to the proteins.
Hence, an improved transfer model weights among GO term associations to a protein for the purpose of more attentively capturing those with more specific GO terms.
Let $\omega(o)$ be a weight is specifically assigned to $o$, the objective of the weighted transfer model is to minimize the following loss,
\begin{equation*}
    \mathcal{L}_{T_2}=\frac{1}{|\mathcal{A}|}\sum_{(o,p)\in \mathcal{A}}\max \left \{ \frac{\omega(o)}{C} \left [ f_{T_2}\left ( o,p \right ) + \gamma^\mathcal{A} - f_{T_2}\left ( o',p' \right ) \right ] , \; 0 \right \}
\end{equation*}
where $C$ is a normalizing constant to constrain that $\sum_{(o,\hat{p})}\frac{\omega(o)}{C}=1$ for a specific protein $\hat{p}$.

\begin{figure}[!ht]
    \centering
    \includegraphics[width=0.8\columnwidth]{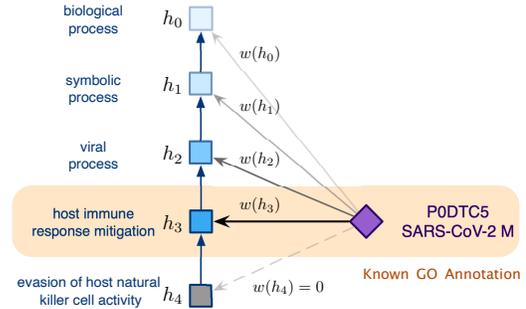}
    \caption{ Explanation of weighted transfer model for modeling hierarchical gene ontology. }
    \label{fig:weighted-transfer}
\end{figure}

Exemplarily, there could be several ways to calculate the association weight.

\stitle{Level-based weight.} The level of the node in one hierarchical taxonomy is a natural indicator of its specificity. Accordingly, the weight can be defined as, 
\begin{equation*}
    \omega(o) = \frac{l}{l_{\max}}
\end{equation*}
where $l$ is the term's current depth and $l_{\max}$ is the maximum length of the associated branch in the gene ontology DAG. 

\stitle{Degree centrality weight.} A small node's degree centrality in the graph roughly reflects its specialty and we apply 
\begin{equation*}
    \omega(o) = \frac{1}{d(o)}
\end{equation*}
as the balance factor for different GO term specialty.

\iffalse
\stitle{GO term specificity weight.} \todo{plan to remove if misleading}
Hyperbolic embeddings can generate by \citep{nickel2017poincare} and the pre-trained embeddings on non-euclidean space reflects the specialty. More specifically, given the $d_h$ hyperbolic embedding of a GO term $\mathbf{h}(o)\in\mathbb{H}^{d_h}$, the specialty weight can be defined as,
\begin{equation*}
    \omega(o) = dist(\mathbf{0},\mathbf{h}(o))
\end{equation*}
where the distance metric is,
\begin{equation*}
    dist(u,v) = \arccosh \left( 1+ \frac{2||u-v||^2}{(1-||u||^2)(1-||v||^2)} \right)
\end{equation*}
From the distance measurement and properties of hyperbolic embeddings, the farther the embedding is from the origin of coordinates, the higher specificity the node represent semantically.
\fi

In practice, incorporating a specificity-based weight to the transfer model essentially enhances the inference in the protein domain, as we have observed in the evaluation in Section \ref{sec:results}.
However, the above weight options generally yield similar performance gain, and we fix the weight option as the level-based weight in our experimental setting.

\subsection{Joint Learning Objectives} \label{subsec:joint}

\modelname jointly learns two knowledge models respectively for GO term relations and PPIs,
and a transfer model to support knowledge transfer between these two.
Therefore, the joint learning objective minimizes the following loss,
\begin{equation*}
    \mathcal{L}= \lambda^t  \mathcal{L}_{T} + \lambda^p \mathcal{L}^{\mathcal{G}_p}_{K} +  \mathcal{L}^{\mathcal{G}_o}_{K}
\end{equation*}
$\lambda^p$ and $\lambda^t$ are two positive hyperparameters.
We use Adam \citep{kingma2015adam} to optimize the learning objective loss.
The learning process uses orthogonal initialization \citep{saxe2013exact} to initialize the weight matrix,
and Xavier normal initialization \citep{glorot2010understanding} for vector parameters.
A normalization constraint is enforced to keep all embedding vectors of GO terms and proteins on unit hyper-spherical surfaces,
which is to prevent the non-convex optimization process from collapsing to a trivial solution where all vectors shrink to zero \citep{bordes2013translating,ma2018hierarchical,yang2014embedding,hao2019joie}.

Note that
\modelname is suitable for joint representation learning on proteomic knowledge of different species. In this protein-GO example, the proteins of these species are significantly different from each other. However, they share the same set of annotations in the GO domain. Therefore, 
More specifically, if we have multiple PPI networks $\mathcal{G}_i, i=1,2,\ldots,m$ where $m$ denotes the number of independent species, $n$ knowledge models are trained respectively. Consequently, one unique transfer model is also trained to facilitate the protein-GO knowledge transfer regarding
each species. The learning objective on the multi-species setting is changed accordingly as,
\begin{equation*}
    \mathcal{L}= \sum^{m}_{i=1} \lambda_i^t \mathcal{L}_{T} + \sum^{m}_{i=1}\lambda_i^p \mathcal{L}^{\mathcal{G}_p}_{K} +  \mathcal{L}^{\mathcal{G}_o}_{K}
\end{equation*}
with the assumption that the knowledge model for gene ontology remains unchanged.

In addition to joint learning on multiple species, \modelname can also be re-trained from new observations of PPIs. For example, suppose newly discovered \covid-Human PPI knowledge extends the original human PPI networks, we can fine-tune the \modelname from the saved model and obtained embeddings, by only optimizing the \modelname on the new triples and hence fast obtain representations for all new proteins, without a long  time for retraining the \modelname from scratch.

%\subsection{Relation Inference} \label{subsec:model_discuss}

% \stitle{Complexity Analysis}
% \junheng{Add complexity analysis if suggested.
\section{Results} \label{sec:results}
In this section, we evaluate {the embeddings learned from} \modelname with two groups of tasks: PPI type prediction (Section \ref{subsec:ppi}) and protein clustering based on {enzymatic functions} (Section \ref{subsec:clustering}).
Furthermore, we provide an extensive case study in Section
\ref{subsec:cov_ppi} on \covid related PPI prediction and classification.

% check each section under the folder "03_Results". 
% (otherwise it is hard to track subsections)
\subsection{Dataset} \label{subsec:dataset}

The protein-protein interactions for yeast (\textit{Saccharomyces cerevisiae}), fly (\textit{Drosophila melanogaster}), and human (\textit{Homo sapiens}) are collected from the STRING~\cite{szklarczyk2016string} database. There are seven types of interactions annotated in the STRING database. To preserve a balanced and sufficient number of cases in each class, we randomly choose the protein pairs from four types of interaction: activation, binding, catalysis, and reaction. In total, there are 21704, 10000, 36400 pairs of proteins for yeast, fly, and human, respectively; each type contains roughly the same number of interactions.
Table~\ref{tab:pro-dataset} summarizes the PPI information for each species. Note that, the human PPI dataset does not contain the virus-generated proteins, but the set partially overlaps with the virus-human pan-PPI networks.

The gene ontology annotations for each protein are extracted from gene ontology Consortium~\cite{gene2018gene}, including all three biological aspects: biological process (BP), cellular components (CC), and molecular function (MF). Table~\ref{tab:go-dataset} summarizes the number of relations between proteins and GO terms. The relations between GO terms include \textit{is-a}, \textit{part-of}, \textit{has-part}, \textit{regulates}, \textit{positively-regulates}, and \textit{negatively-regulates}.

%Protein interactions and annotations are collected from UniProt~\citep{uniprot2019uniprot} and the gene ontology is extracted from AMIGO2~\citep{gene2018gene}. 
%The statistics of extracted proteins (yeast, fly, and human\footnote{This human PPI dataset only include general human protein interactions that were previously established before the COVID-19 outbreak. That is, it excludes virus-generate proteins but partially overlaps with the virus-human pan-PPI networks.}) is listed in Table \ref{tab:pro-dataset}. Data statistics of the gene ontology is listed in detail in Table \ref{tab:go-dataset} with its three major sub-domains: biological process (BP), cellular components (CC) and molecular functions (MF) respectively.  

\begin{table}[htbp]
\centering
\caption{Statistics of PPI networks and associated GO annotations from different species. }
\vspace{-8pt}
% \resizebox{\linewidth}{!}{
%\setlength\tabcolsep{3pt}{
%\small
\begin{tabular}{l|c|c|c}
\newtoprule
\textbf{Species} & \textbf{\# Proteins}  & \textbf{\# PPI Triples} & \textbf{\# GO Annotations}  \\
\newmidrule
Yeast   & 3,736  & 21,704 & 191,801 \\
Fly     & 3,826  & 10,000 & 87,807\\
Human   & 8,204  & 36,400 & 102,759 \\
\newbottomrule
\end{tabular}
%}
\label{tab:pro-dataset}
\vspace{-4pt}
\end{table}
\begin{table}[htbp]
\centering
\caption{Statistics of three aspects in the gene ontology: biological processes (BP), cellular components (CC) and molecular functions (MF).}
\vspace{-8pt}
%\resizebox{\linewidth}{!}{
%\setlength\tabcolsep{3pt}{
%\small
\begin{tabular}{l|c|c|c}
\newtoprule
\textbf{Aspects} & \textbf{BP}  & \textbf{CC} & \textbf{MF} \\
\newmidrule
\# GO entities   & 5744     & 1,147 & 1,764\\
\# GO triples    & 19,021   & 2,116 & 2,190\\
\newmidrule
\# Protein-GO annotations (yeast) & 72,956 & 58,729 & 60,116 \\
\# Protein-GO annotations (fly)   & 44,605 & 24,550 & 18,652 \\
\# Protein-GO annotations (human) & 42,899 & 32,929 & 26,931 \\
\newbottomrule
\end{tabular}
%}
\label{tab:go-dataset}
\vspace{-8pt} %space control
\end{table}

For the \covid dataset, we collect the latest virus-protein interaction from BioGrid\footnote{Data source: \href{https://wiki.thebiogrid.org/doku.php/covid}{https://wiki.thebiogrid.org/doku.php/covid}} and the limited GO annotations for \covid from Gene Ontology Consortium\footnote{Data source: \href{http://geneontology.org/covid-19.html}{http://geneontology.org/covid-19.html}}, as last updated on early April.
In summary, there are 26 \covid generated proteins and 332 human proteins presenting the evidence of viral-human protein interactions as suggested by ~\citet{gordon2020sars}. The selection is based on a high MIST score and a low SAINTexpress BFDR from Affinity Capture-MS. Out of the same experiment, we select 1131 viral-human protein pairs with MIST scores lower than 0.01 as our negative samples. The 26 \covid generated proteins are annotated with 282 GO terms.
In addition to \covid, BioGrid also includes 30 viral proteins from SARS-CoV and MERS-CoV, which are two similar contagious viruses causing respiratory infection. 
These 30 viral proteins are annotated with 630 GO terms, and display 326 interactions with human proteins.
All processed datasets are available at \href{https://www.haojunheng.com/project/goterm}{https://www.haojunheng.com/project/goterm}.

% delete subjective assumptions on dataset introduction
% \chelsea{need number here to support that they are sparse} \junheng{Will add some sparsity metric: Yeast has only roughly 0.6\% of all possible pairs of proteins that are known to interact.}
% It is worthy to note that from the statistics listed above, we confirm that all the collected yeast PPI, fly PPI, human PPI, and even \covid-human pan-PPI, the interaction graphs are generally sparse in terms of the average number of transactions per proteins. 

\subsection{Baselines} \label{subsec:baselines}
%We compare \modelname with the following state-of-the-art approaches on learning representation on proteins (and gene ontology, if applicable).
We compare \modelname with the most applicable state-of-the-art approach, Onto2Vec~\cite{smaili2018onto2vec}, on learning the representation of proteins. Onto2Vec considered the annotation from gene ontology for representation learning. In addition, we compare \modelname with a simpler setting, \modelname-NonGO, where we only consider the single-domain knowledge of PPI.
%Various settings on hyperparameters are considered for some of the baselines. \todo{Still mention binary?}

% \stitle{Binary} Proteins are simply encoded by vectors of binary elements, which come from \chelsea{one-hot embedding source (by Chelsea)}. \CJ{Binary is not recommended for reporting.}

\stitle{Onto2Vec, Onto2Vec-Parent, Onto2Vec-Ancestor}. Onto2Vec utilizes the annotation information from gene ontology to create pairwise context and apply Word2Vec \citep{mikolov2013distributed} to generate protein and GO term embeddings. Its schema allows the model to learn the representation of proteins and GO terms simultaneously. The proposed setting of Onto2Vec only includes the direct relationship between a protein and a GO term. In this experiment, we explicitly include the relationship between a protein and the parents of the annotated GO terms, named \textit{Onto2Vec-Parent}, and the ancestors of the annotated GO terms, named \textit{Onto2Vec-Ancestor}. 

%We select four variants noted as \emph{Parent} (using partial GO which is linked to proteins with their parent GO term), \emph{Ancestor} (using the entire GO), \emph{Sum} (protein embeddings computed from the sum of GO embeddings) and \emph{Mean} (protein embeddings computed from the average of GO embeddings). 
%\junheng{explain the meaning of all variants.}
\stitle{Onto2Vec-Sum, Onto2Vec-Mean}. To examine the effect of Onto2Vec on learning the protein representation from a single domain, i.e. gene ontology, we remove the relations between proteins and GO terms during the learning process. The representation of a protein is then computed by either summing up the embeddings of all the associated GO terms (\textit{Onto2Vec-Sum}), or taking the average of the embeddings of those GO terms (\textit{Onto2Vec-Mean}).

\stitle{OPA2Vec} Based on Onto2Vec, OPA2Vec further learns the protein and GO term embeddings by leveraging meta-data (labels, synonyms, etc), which better characterize GO terms.

\stitle{\modelname (NonGO)}. As opposed to considering the knowledge from a single domain of gene ontology, we adopt \modelname to consider only the knowledge from Protein-Protein Interaction. 
In this approach, all the gene ontology annotations and the gene ontology graph are neglected, and thus is reduced to a knowledge model. We only use the knowledge model in Section \ref{subsec:knowledge_model}, where the protein embeddings are solely learned from PPI networks by the original KG embedding technique, DistMult. We refer to this approach as ``Non-GO''. 
% \chelsea{What does KM stand for?}
%That is, ``KM-Only'' setting means the PPI prediction are only made based on the single-domain knowledge.

It is worth mentioning that the goal of Onto2Vec and OPA2Vec is to learn the protein representation; therefore, to adapt for the task of PPI prediction, we concatenate the embeddings of each pair of proteins and train a multi-class classifier to predict the PPI type for a given pair of query proteins. We examine the performance with four different classifiers: logistic regression (LR), support vector machine (SVM), random forest (RF), and neural networks (MLP). The evaluation is conducted with five-fold cross-validation. Similar settings apply to all Onto2Vec variants and OPA2Vec. On the contrary, 
%It is noteworthy that existing methods generally follow the pipeline which first obtains the protein feature-based embedding at first and later trains a multi-class classifier to predict the PPI type given pairs of query proteins.
%This is a contrast to 
our proposed model equips with relational modeling and outputs PPI predictions by selecting the most plausible relation type. As a result, we do not need an additional classifier for \modelname and \modelname-NonGO. 
%\chelsea{Did we try cross-validation?} \junheng{Mentioned in 3.3 experimental setting}

%For these baselines that require a statistical classification model, we adopt logistic regression (LR), support vector machine (SVM), random forest (RF), and neural networks (MLP), as typical options. Therefore, ``Onto2Vec+SVM'' denotes that the Onto2Vec GO term embeddings are used as features to train an SVM for predicting PPI types.
%In Section \ref{subsec:ppi} and \ref{subsec:clustering}, due to limited space, we only report the results from the statistical classification baseline with the best performance. % dataset and baselines

\subsection{PPI Type Prediction on Multiple Species} \label{subsec:ppi}
%\todo{whether to change ``multiple'' in section title.}
We examine how effectively \modelname leverages gene ontology to predict protein-protein interaction types. 
To do so, we first evaluate the performance on three organisms separately: human, yeast, and fly. 
Then we study the contribution of the three aspects in gene ontology, i.e. biological process (BP), cellular component (CC), and molecular function (MF), on predicting the type of PPI. 
%In addition to joint learning on two different domains (i.e. GO terms and PPI), we also investigate the power of joint learning from multiple species. \chelsea{need more details here?} \junheng{``QX'' goals with questions deleted.}
Specifically, we provide an analysis on how the knowledge from Gene Ontology contributes to PPIs in different species.
%\junheng{Do we still mention the benefit of joint training on multiple species?}
% \stitle{Goals} In this subsection, we try to answer the following questions, based on the experiments on PPI type prediction: 
% (1) Q1: How \modelname can better leverage the complementary knowledge in the gene ontology domain and improve the performance on PPI type prediction, compared to state-of-the-art baselines;
% (2) Q2: If incorporating the knowledge from GO is beneficial, which subdomain(s) in GO (i.e. BP, CC, MF) are more informative and crucial for PPI type prediction?
% (3) Q3: How \modelname can adapt to integrate the knowledge  from multiple species and potentially further improve PPI type prediction?

\stitle{Experimental setting.} 
%We  evaluate our \modelname model on the PPI prediction task for three species, i.e. yeast, fly, and human. 
%For training, we input the PPI training set, GO annotations of the proteins in the training folds and the entire gene ontology graph. 
% \chelsea{need to explain how we set up the training, not just the input for training. how do we obtain the best performing hyperparameters?} \junheng{added}
We first separate the PPI triples into  approximately 70\% for training, 10\% for validation and 20\% for testing.
For hyperparameters with the best performance from the validation set, we select dimension $d_p = d_o = 300$ and margin parameters $\gamma^{\mathcal{G}}=0.25$, $\gamma^{\mathcal{A}}=1.0$ and $\gamma_{\text{HA}}=1.0$. Two weight factors in the joint learning objective are set as $\lambda_p=1.0, \lambda_t=1.0$. We use DistMult for the knowledge model in Section \ref{subsec:knowledge_model}, with 
hierarchy-aware regularization and the level-weighted transfer model (Section \ref{subsec:transfer_model}) deployed. For simplicity, the reported \modelname adopt the same settings if not specifically explained. The number of epochs in training on all settings is limited to 150. 
%We select different input sources of protein interaction networks and gene ontology based on the aforementioned goals and change the setting accordingly.
For evaluation, we aim at predicting the correct interaction type, given pairs of proteins in the test set.
We conduct a 5-fold cross validation for \modelname and all baselines, and report the average and standard deviation of accuracy. 
%\todo{GO ``parent'' setting}
% \chelsea{standard deviation is still missing from the table.}
% \chelsea{the classifier for Onto2Vec?} 
%As for Onto2Vec and OPA2Vec, we select the best performing classifier mostly as RF. One exception is that we apply MLP for Onto2Vec-Ancestor on fly.
The best-performing classifier is RF for OPA2Vec and most of the Onto2Vec variants. The only exception is to apply MLP for Onto2Vec-Ancestor on fly.

%%%%%%%%%%%%%%%%%%%%%%%%%%%%%%%%%%%%%%%%%%%%%%%%%%%%%%
% PPI Part 1: Baseline comparison
%%%%%%%%%%%%%%%%%%%%%%%%%%%%%%%%%%%%%%%%%%%%%%%%%%%%%%
\begin{table}[htbp]
\centering
\caption{ PPI type prediction accuracy (\%) evaluated on yeast, fly and human species.}
\vspace{-4pt}
\resizebox{\linewidth}{!}{
\small
\begin{tabular}{l|c|c|c}
\newtoprule
\textbf{Model} & \textbf{Yeast} & \textbf{Fly}  & \textbf{Human} \\
\newmidrule
% Binary  			&  7896 &  7530 &  6743\\
Onto2Vec             &  76.41 $\pm$ 0.73 &  70.85 $\pm$ 0.85 &  77.97 $\pm$ 0.46\\
Onto2Vec-Parent    &  80.79 $\pm$ 0.66 &  75.46 $\pm$ 1.11 &  74.90 $\pm$ 0.46\\
Onto2Vec-Ancestor  &  86.31 $\pm$ 0.42 &  80.31 $\pm$ 0.92 &  78.73 $\pm$ 0.46\\
Onto2Vec-Sum  	 &  76.38 $\pm$ 0.83 &  72.84 $\pm$ 1.13 &  72.53 $\pm$ 0.73\\
Onto2Vec-Mean      &  77.95 $\pm$ 0.81 &  74.38 $\pm$ 1.13 &  73.47 $\pm$ 0.80\\
OPA2Vec              &  79.88 $\pm$ 0.74 &  74.45 $\pm$ 0.97 &  72.04 $\pm$ 0.58\\
\newmidrule
\rowcolor{shadecolor} \modelname-NonGO   &  83.65 $\pm$ 0.92     &   77.58 $\pm$ 1.07 &  76.10 $\pm$ 0.87\\
\rowcolor{shadecolor} \modelname    &  87.15 $\pm$ 1.15 &   84.56 $\pm$ 0.81 &  81.42 $\pm$ 0.62 \\
\rowcolor{shadecolor} \modelname-Weighted & \textbf{ 90.12 $\pm$ 1.21} & \textbf{ 85.55 $\pm$ 1.57} & \textbf{ 83.89 $\pm$ 0.92}\\
\newbottomrule
\end{tabular}
}
\label{tab:ppi}
\vspace{-6pt}
\end{table}

\stitle{Results.} % First, we train \modelname with all Gene Ontology annotations with PPIs in train set on three organisms separately. \chelsea{Need to fix this sentence.}
%: human, yeast, and fly. 
The results for PPI type prediction are shown in Table \ref{tab:ppi}. 
We observe that our best \modelname variant outperforms \modelname-NonGO by {7.4\%} on average for all three species. This observation directly shows that gene ontology KG provides complementary knowledge for proteins. Subsequently, Gene Ontology annotations benefit the learning of protein representations and better predict the interaction types between proteins. 
%\chelsea{What kind of evidence showing that the knowledge from gene ontology is ``complementary''?}\junheng{we changed to ``augmented'' globally.}
Compared to other baselines, it is observed that \modelname notably outperforms Onto2Vec-Ancestor with an average increase of 7.4\% on the prediction accuracy, and a relative gain of 9.0\% on average of all three species. This observation is due to the advantage that \modelname better leverages the complementary knowledge from PPI to enhance the PPI prediction. As mentioned in Section \ref{subsec:baselines}, Onto2Vec does not utilize the PPI information into protein embedding learning. Instead, it obtains embeddings based on the aggregated semantic representations of GO terms. It requires additional classifiers for PPI type prediction given pre-trained protein embeddings. In contrast, \modelname jointly learns protein representations from both the knowledge model that captures the structured information of known PPIs, and the transfer model that delivers the annotations of GO terms. Also, we observe that \modelname-Weighted achieves better results than \modelname, with a relative performance gain of {2.5\%}. We hypothesize that such gain is attributed to specificity modeling in the transfer model which distinguishes more specific and informative GO terms from other general GO terms and assigns a higher weight, which selectively learns the alignments between two domains.
In terms of different species, we also observe that \modelname achieves a higher PPI prediction accuracy on yeast compared to human and fly. The possible reason is that the yeast interaction network is denser, such that {0.30\%} of the protein pairs are known to interact, compared to human ({0.13\%}) and fly ({0.11\%}), which indicates that yeast is possibly well studied. 
%\todo{check explanation. Trying to explain it by the density of PPIs in OUR dataest stats.}
OPA2Vec claims to be an improved version of Onto2Vec. Similar to Onto2Vec, it only considers the direct relationship between a protein and a GO term, without parents and ancestors. We find that OPA2Vec performs slightly better than Onto2Vec on Yeast and Fly, but worse on Human. 
In addition, OPA2Vec falls short when compared to any of the \modelname variants, indicating that incorporating the metadata of GO terms is insufficient for protein representation learning.

%\junheng{We find that OPA2Vec does not outperform the strongest Onto2Vec variant because OPA2Vec simply includes metadata of GO terms which does not significantly benefit protein representation learning.}

It is noteworthy that unlike Onto2Vec, which achieves its best performance with the help of full gene ontology (i.e. Onto2Vec-Ancestor),  our \modelname model can utilize only the GO terms that are directly annotated with the proteins to accomplish the highest accuracy score. This also makes \modelname training processes more time efficient.
% \todo{We never mention that our GO terms setting includes the parents.} \junheng{Current data is ``children'' setting?}
% More specifically, the results on fly PPI type prediction show that \modelname trained with direct gene ontology annotations of proteins has an accuracy of 85.55\%, while \modelname trained with all ancestors of annotated GO terms reaches a suboptimal performance of 83.90\% on accuracy.
%\chelsea{Where are the data?}\junheng{Not listed in the figure though. Just mention in text.}
We hypothesize that for \modelname in the PPI type prediction task, {GO terms that are directly related to associated proteins with high specificity are sufficient for the transfer model to model the protein-GO association in the embedding spaces.} %\junheng{did not mention why add more GO term ancestor can make it worse.}
In contrast, Onto2Vec needs entire structured information of GO terms for its word2vec module to construct an exhaustive context of protein features.

%%%%%%%%%%%%%%%%%%%%%%%%%%%%%%%%%%%%%%%%%%%%%%%%%%%%%%
% PPI Part 2: Three different categories of GO
%%%%%%%%%%%%%%%%%%%%%%%%%%%%%%%%%%%%%%%%%%%%%%%%%%%%%%
\begin{table}[htbp]
\vspace{-4pt}
\centering
\caption{Comparison of PPI prediction accuracy of \modelname on three different aspects of gene ontology.}
\vspace{-4pt}
%\resizebox{\linewidth}{!}{
%\setlength\tabcolsep{3pt}{
%\small
\begin{tabular}{c|c|c|c|c}
\newtoprule
\textbf{\#} & \textbf{Aspects}  & \textbf{Yeast} & \textbf{Fly} & \textbf{Human} \\
\newmidrule
\multirow{3}{*}{ \tabincell{c}{1}} & BP  & 0.8794 & 0.8402 &  0.8153\\
&  CC	    & 0.8499 & 0.8272 & 0.8054\\
&  MF	    & 0.8539 & 0.8386 & 0.8165\\
\newmidrule
\multirow{3}{*}{ \tabincell{c}{2}}  & BP+CC	& 0.8717 & 0.8473 & 0.8271 \\
& BP+MF	& 0.8673 & 0.8471 & 0.8163\\
& CC+MF	& 0.8569 & 0.8466 & 0.8170\\
\newmidrule
3 &  AllGO	& \textbf{0.9012} & \textbf{0.8555} & {\textbf{0.8389}} \\
\newbottomrule
\end{tabular}
%}
\label{tab:ppi-threeview}
\vspace{-4pt}
\end{table}
%We further explore the effects of subdomains of gene ontology that is leveraged for PPI type prediction. 
We further explore the effects of three different aspects of gene ontology in predicting the types of PPIs. 
To achieve this, %we construct the cross-domain KB with specific categories of GO terms and train \modelname with different combinations of GO categories.
we train \modelname in settings where only specific aspects of gene ontology annotations are used.
Results are shown in Table \ref{tab:ppi-threeview}, in which %BP refers to biological process, CC refers to cellular component, and MF refers to molecular function. 
BP, CC and MF respectively refer to the cases where GO terms of \emph{biological processes}, \emph{cellular components} and \emph{molecular functions} are used.
``BP + CC'' denotes that the GO terms from both biological processes and cellular components are included in training.
%That is,  ``BP+CC'' denotes that only GO terms that are biological process or cellular components are used to train \modelname.  
%Based on the observation that \modelname with full three GO domains (BP+CC+MF) outperform \modelname with two subdomains (BP+CC. BP+MF, CC+MF), which also dominate that with only one subdomain (BP, CC, MF), we hypothesize that all three views can improve the protein representation learning and positively affect the PPI accuracy. 
We observe that \modelname performs the best with GO terms from all aspects (full gene ontology).
This phenomenon is consistent among all three species, indicating that the protein representations are more robust when learning from a more enhanced knowledge graph.
%Also, comparing the average performance gain by adding one subdomain in Table \ref{tab:ppi-threeview}, biological processes has better improvement (around 2.61\%)when added than cellular components (around 2.42\%) and molecular functions (around 2.13\%), because annotations on biological processes contain much more entities and relational facts and form a larger subdomain while annotations on cellular components and molecular functions can efficiently improve the PPI performance with a limited number of GO entities. 
It is also interesting to see that the accuracy of the task is generally higher when we include the GO terms from biological processes. This leads to 2.61\% improvement in accuracy over CC, and at least 2.13\% of improvement over MF when evaluating individually. In the two-aspect evaluation, ``BP+CC'' is in average leads to 0.7\% better accuracy than ``CC+MF''. 
%\junheng{Since the number of percentage is small, use relative percentage gain instead of fold-change expression.}
%\todo{List out these numbers in the table; otherwise, it is better to leave them out of the text}.
This is attributed to the fact that BP is the largest group in the gene ontology, containing more entities and relational facts. 
Consequently, \modelname achieves the best performance with all three aspects of gene ontology annotations incorporated. 
This indicates that the characterization of PPIs benefits from more comprehensive gene ontology annotations.

%This indicates that rich annotations indeed can help learn protein embeddings by applying the transfer model. 
% \junheng{It is noteworthy that for different species, the performance gain from multi-species joint training varies when incorporating a new subdomain of gene ontology. This effect is more obvious in the PPI prediction on \covid, as explained later in Section \ref{subsec:cov_ppi}}
% \chelsea{Move this claim to the covid section} 

%%%%%%%%%%%%%%%%%%%%%%%%%%%%%%%%%%%%%%%%%%%%%%%%%%%%%%
% PPI Part 3:  Benefit of joint training
%%%%%%%%%%%%%%%%%%%%%%%%%%%%%%%%%%%%%%%%%%%%%%%%%%%%%%
\begin{table}[htbp]
\centering
\caption{PPI type prediction accuracy on different configurations of multi-species joint learning.}
\vspace{-4pt}
%\resizebox{\linewidth}{!}{
%\setlength\tabcolsep{3pt}{
%\small
\begin{tabular}{l|c|c|c}
\newtoprule
\textbf{Model} & \textbf{Yeast} & \textbf{Fly} & \textbf{Human} \\
\newmidrule
\modelname(single)      & 0.9012 & 0.8555 & {0.8389}\\
\modelname(concat)       & 0.8795 & 0.8282 & {0.8028} \\
\modelname(multi-way)    & \textbf{0.9062} & \textbf{0.8638} & \textbf{0.8426} \\
\newbottomrule
\end{tabular}
%}
\label{tab:ppi-multi}
\end{table}
% \chelsea{@Muhao, can you help with the following paragraph?}
% \junheng{Discuss: The name of three settings on multi-species joint learn: ``single'', ``concat'', ``multi-way'' or ``joint''?}
In addition to joint learning from two different domains (i.e. GO terms and PPIs), as mentioned in Section \ref{subsec:joint}, \modelname 
%has the ability to fuse the embeddings for multiple species with species-specific knowledge models and the transfer models, together with the universal species-independent gene ontology. 
can be trained to capture PPIs for multiple species with several species-specific knowledge models, along with transfer models that bridge the universal gene ontology.
To validate the benefit of joint learning on multiple species together, we consider three following configurations of \modelname: (i) the ``multi-way'' setting uses one unique knowledge model and one transfer model to the universal gene ontology for each species; (ii) the ``concat'' setting uses one unified knowledge model to capture all species of PPIs, together with one transfer model to learn protein-GO alignments, that is, simply concatenate all PPI triples and all gene ontology annotations of proteins in multiple species;
(iii) the ``single'' setting trains separately on each species, which is exactly the same as in the setting in Table \ref{tab:ppi}.
We summarize the results in Table \ref{tab:ppi-multi}. It is observed that the ``multi-way'' setting can slightly improve PPI performance in comparison to the `` single'' setting that trains separately on each species. Also in the ``concat'' setting with one shared transfer model and knowledge model, the performance significantly drops with a 2.8\% decrease of accuracy on average compared to the ``single'' setting. Such results suggest that each species has unique patterns of PPIs, %and sharing the same knowledge and transfer model will compromise the ability to learn accurate protein representations while our designated \modelname from multiple species can mutually enhance the PPI performance.
%Therefore, mixing all species and using one transfer model may introduce inconsistency to the model learning. 
%This is attributed to that different species may have different species-specific features and annotations and encoding all protein entities
such differences are better differentiated in separate embedding spaces.
Hence, the multi-way setting better encodes the species-specific knowledge and model, which helps the type prediction of PPIs for each species by \modelname that are jointly trained on multiple species. 

%In summary, we demonstrate that \modelname can better incorporate the complementary knowledge from gene ontology for \todo{for what?}
%\chelsea{We can only conclude about the effect for PPI task here}\junheng{add a summary if space allowed}
\subsection{Identifying Protein Families And Enzyme Commission Based Clustering } \label{subsec:clustering}

%\modelname representations for proteins not only can be used for PPI type prediction as one relational inference task, but also can help identify potential protein families based on functional annotation. 
Besides inferring PPI types, the embedding representations of proteins can also be used to identify potential protein families based on their functions.
%\chelsea{protein family is a dangerous term, usually it is categorized by structure. I recommend to be more specific here.}
This can be achieved by performing clustering algorithms on the learned protein embeddings.

\begin{comment}
%For the protein clustering task, we use all available information for training the protein embeddings. 
%More specifically, we use the entire PPI network as input, together with all GO annotations of proteins and the same GO graph in Section \ref{subsec:ppi}. 
We learn the protein representations using the knowledge from the PPI network and from gene ontology.  
We also include two different settings, named transductive and inductive. For the transductive setting, %all proteins are observed in the PPI network, i.e. all proteins have at least one known interaction type with other proteins.
we only include the proteins that are observed in the PPI network.
%In contrast, in the inductive setting, 
For the inductive setting, we include all of the proteins that are annotated with GO terms.
Some proteins only have gene ontology annotations but do not have any known PPT triples with other proteins, which are considered as ``low-resource proteins''. 
All other hyperparameters remain the same as Section \ref{subsec:ppi}. 
\chelsea{Why name them transductive and inductive? What is the purpose of doing so?} \junheng{In transductive setting, proteins without GoTerm alignment will be ignored which makes proteins with embeddings fewer.}

\chelsea{For this experiment, do we only consider the balance dataset for PPI or all of the proteins? If using only a subset of the proteins, is it still fair?} \junheng{will change to report inductive setting only.}
\end{comment}

The Enzyme Commission number (EC number) defines a hierarchical classification scheme that provides the enzyme nomenclature based on enzyme-catalyzed reactions. The top-level EC numbers contain seven classes: oxidoreductases, transferases, hydrolases, lyases, isomerases, ligases, and translocases. 
In this experiment, we select 1340 yeast proteins in total with enzymatic functions.
We learn the protein representations using all the triples of PPI networks and the annotation from gene ontology and evaluate the learned representations of these proteins by performing the k-means clustering algorithm to group them into seven non-overlapping clusters. These clusters are compared with the top-level of enzyme commission classification.
%For evaluation, given the obtained protein embeddings, we perform K-means clustering method and compared the clustering results with hand-labeled 1340 proteins on yeast based on their EC categorization\footnote{For transductive setting, since not all protein in the labeled proteins are observed in the PPI network, we have slightly fewer available protein embedding, which are 1066 proteins on yeast.}.
Purity score is reported as evaluation metrics. 

%\stitle{Results.}
The evaluation of the clustering results is shown in Table \ref{tab:ec_cluster}. 
\modelname achieves the best clustering performance on yeast by a relative increase of {9.7\%}, which demonstrates that \modelname has the good model capability to representation learning and empirically show the validity of the learned embeddings to measure the similarity.
%\chelsea{please put the the relative improvement in the table, otherwise, readers will not know where this number comes from}. \todo{add relative gain in table.}
We hypothesize that \modelname better incorporates protein annotation resource and utilizes the complementary knowledge in the gene ontology domain, while \modelname also captures PPI information and encode it into protein embeddings. This in the end results in comprehensive representations for proteins and helps to identify protein EC classes by clustering.

% \chelsea{what does it mean by ```especially by incorporating known PPI'''? Are you trying to say that \modelname is better because it incorporates more resources?}

% In the inductive setting, \modelname outperform most of Onto2Vec variants. 
% Comparing two different settings, we observe that for both Onto2Vec and \modelname, there is a significant drop in clustering performance since the low-resource proteins which lack PPI information. 

% \chelsea{The experiment needs a new design. It seems like the proteins used in transductive is a subset used in inductive. It is not an apple-to-apple comparison when discuss the differences of transductive vs inductive. I have two questions here: 1) are we trying to emphasize the performance of those low-resource proteins? If so, we should only look at their performance.  2) are we trying to emphasize the benefit of including these low resource proteins? if so, we should look at the performance of other proteins.}
% \junheng{new version delete transductive and only inductive.}
% \chelsea{We also need to explain why the ARI and Purity are very low.}\junheng{The purity score compared to Onto2Vec are comparable. One solution is to only provide purity score without NMI. It seems that \modelname is better than Onto2Vec on purity but both are not good. My suggestion is to shorten this subsection without mentioning low-resource proteins.}

\begin{table}[htbp]
\centering
\caption{Results of top-level EC clustering by K-means on learning selected yeast protein embeddings.}
\vspace{-8pt}
%\resizebox{\linewidth}{!}{
%\setlength\tabcolsep{3pt}{
%\small
\begin{tabular}{c|c}
\newtoprule
\textbf{Model} & \textbf{Purity Score} \\
\newmidrule
Onto2Vec              & 0.2339 \\
Onto2Vec-Parent       & 0.2452 \\
Onto2Vec-Ancestor     & 0.3224 \\
Onto2Vec-Sum  	    & 0.3022 \\
Onto2Vec-Mean         & 0.2616 \\
\modelname (KM only)    & 0.2514 \\
\newmidrule
\modelname			    & \textbf{0.3306} \\
% \multirow{2}{*}{ \tabincell{c}{\textbf{\modelname}}} & \textbf{0.0507} & \textbf{0.3306} \\
% & \footnotesize{(+0.030)} & \footnotesize{(+0.008)} \\
\newbottomrule
\end{tabular}
%}
\label{tab:ec_cluster}
\vspace{-8pt}
\end{table}

\subsection{Case Study: SARS-CoV-2-Human Protein Target Prediction} \label{subsec:cov_ppi}

%\junheng{Section title check.}
The COVID-19 pandemic requires many efforts and attentions from scientists of different fields. However, there is very limited knowledge of the molecular details of \covid.
In this subsection, we apply \modelname to gain more insights of the PPI network between \covid and human proteins.
Specifically, we explore the potential of \modelname on predicting whether a pair of human and \covid proteins interact or not. This is modeled as a binary prediction task. 
Correspondingly, results from the binary predictions can serve as a guide to identify the targeted proteins by \covid. 
We first use the known interactions between these two species to validate the effectiveness of \modelname. These interactions are experimentally verified as described in Section~\ref{subsec:dataset}. 
%In this subsection, we apply the \modelname to predict \covid-Human related PPIs. 
%We explore how \modelname performs given partially known \covid interaction information with human protein targets and \covid related GO term annotations from the gene ontology.
In this setting, we particularly study the contribution of the knowledge of other closely related viruses (SARS-CoV and MERS) on supporting PPI prediction.
We also show the high-confidence candidates of targeted human proteins predicted by \modelname for four selected \covid proteins.  
%We also show how the proposed model can help provide high-confidence candidates for potential human protein targets. 

\stitle{Experimental setting.} %Unlike PPI type prediction that seek to predict the most possible interaction type for interacting proteins, a more realistic and useful task 
%Because we have limited resources of \covid human-virus PPIs, \todo{reason} is to identify whether a pair of one virus-generated protein and one existing human protein would interact with each other, as a binary classification task.
In this experiment, we randomly split the known positive human-virus PPIs into train and test sets with a ratio of 80\% and 20\%. We train \modelname on this train set along with human PPIs.
For evaluation, positive test samples and selected negative samples, mentioned in Section \ref{subsec:dataset} are used to perform binary prediction. We adopt F1-score as the evaluation metric.

\begin{figure}[!ht]
    \centering
    \includegraphics[width=0.7\columnwidth]{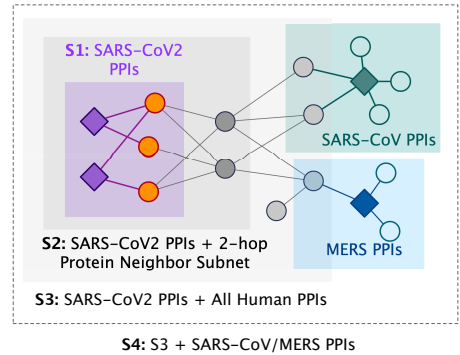}
    \vspace{-4pt}
    \caption{Different scopes of input to train \modelname for \covid PPI prediction. }
    \label{fig:subset_def}
\end{figure}
\vspace{-8pt}

\begin{table}[htbp]
\centering
\caption{F-1 score on \covid-Human PPI interaction classification.}
\vspace{-6pt}
%\resizebox{\linewidth}{!}{
%\setlength\tabcolsep{3pt}{
%\small
\begin{tabular}{c|c|c|c|c}
\newtoprule
{Input} &  S1 & S2 & S3 & S4 \\
\newtoprule
NonGO       &0.6737	 &0.7004	 &0.6918	 &0.6997\\
\newmidrule
BP           &0.7103	 &0.7353	 &0.7348	 &0.7492 \\
CC           &0.7188	 &0.7383	 &0.7380	 &0.7675 \\
MF           &0.6737	 &0.7016	 &0.7022	 &0.7365 \\
\newmidrule
BP+CC        &0.7257	 &0.7570	 &0.7499	 &0.7813\\
BP+MF        &0.7252	 &0.7479	 &0.7486	 &0.7713\\
\rowcolor{shadecolor}  
CC+MF        &0.7317	 &0.7622	 &0.7692	 &\textbf{0.7917} \\
\newmidrule
AllGO       &0.7307	 &0.7537	 &0.7500	 &0.7885\\
\newbottomrule
\end{tabular}
%}
\label{tab:covid-binary}
\vspace{-8pt}
\end{table}

% As for the classification task, the results are in Table \ref{tab:covid-binary}. \todo{Results in SubNet selection and view selection. Conclusion: CC+MF are most useful.}
% \todo{Indeed it is generally not a very strong classifier to test set (~50 test pairs). We may select to report other metrics or report on all positive and negative samples, which is not standard.}

\stitle{Results.}  As in Section \ref{subsec:ppi}, we first evaluate \modelname on \covid PPI prediction. From the observation in Section \ref{subsec:ppi}, two important factors are considered: three aspects in the gene ontology domain and the scope of input \covid-Human PPIs. More specifically, we define increasingly four scopes of input PPIs, as shown in Figure \ref{fig:subset_def}, i.e. (1) S1: Only using the train folds of \covid-Human PPIs; (2) S2: Using \covid-Human PPIs with the 2-hop neighbor proteins from \covid viral proteins, i.e. including the ones which also interact with any proteins that the \covid interacts; (3) S3: \covid-Human PPIs with all other protein interactions on human; (4) S4: \covid-Human PPIs with all protein interactions in S3 plus all SARS-CoV and MERS PPIs. 
As for the aspects of the gene ontology domain, similar to Table \ref{tab:ppi-threeview} in Section \ref{subsec:ppi}, we adopt eight options, i.e. one without gene ontology information (NonGO), three using a single aspect of GO terms (BF, CC, MF), three options using two of the aspects (BF+CC, etc) and one using all three aspects (AllGO).

The results are summarized in Table \ref{tab:covid-binary}. In terms of gene ontology aspects, we observe that CC contributes the most compared to other aspects of gene ontology annotations, and the best performance is achieved by adopting CC+MF in \modelname learning. One explanation is that most of the \covid proteins have CC annotations and these annotations make up over 70\% of all currently available annotations on average. However, less than 5 proteins (such as NSP and ORF 1a) have BF and MF annotations, 
possibly due to insufficient knowledge on understanding \covid biological mechanism. 
As for the input fields, we find that the performance drastically increases with the expansion of input from S1 to S2, which indicates that interactions of 2-hop neighbor proteins can benefit \covid PPI prediction. However, such a trend is not clearly observed when expanding the input field from S2 to S3. We hypothesize that proteins that are not within 2-hop neighbors may not be very related to \covid or provide beneficial insights. 
Interestingly, when adding interactions of two related coronaviruses (SARS-CoV/MERS-CoV) that cause respiratory infection, the performance continues to improve with a relative gain of 3.4\%. As shown in Figure. \ref{fig:covid-go-example}, 
%similar viruses in the past are closely related to \covid and share important properties.
viruses that are closely related to \covid tend to share important properties. 
This strongly suggests that it is crucial to leverage their interactions and gene ontology annotations as augmented knowledge for drastically emerging \covid.

\begin{table}[htbp]
\centering
\caption{Top target proteins predicted by \modelname. Known interactions from training set are excluded. Proteins that are considered as high-confidence targets are boldfaced. }
\vspace{-8pt}
%\resizebox{\linewidth}{!}{
%\setlength\tabcolsep{3pt}{
%\small
\begin{tabular}{c|p{6cm}}
\newtoprule
\textbf{\covid} & \textbf{Targeted proteins in human}\\
\newmidrule
ORF8 & 
% \textbf{P05556\footnotesize(0.817)}, 
% \textbf{Q9Y4L1\footnotesize(0.772)}, 
% P17858{\footnotesize(0.393)}, 
% \textbf{Q9NXK8\footnotesize(0.824)}, 
% Q9BQE3{\footnotesize(0.310)}
\textbf{P05556}, P61019, \textbf{Q9Y4L1}, P17858, Q92769,  Q9BQE3, Q9NQC3,  \textbf{Q9NXK8}, P33527, P61106
\\
\newmidrule
NSP13 &
% \textbf{Q99996\footnotesize(0.990)},
% \textbf{P35241\footnotesize(0.912)},
% Q12923{\footnotesize(0.587)},
% Q86YT6{\footnotesize(0.353)},
% \textbf{Q9Y2I6\footnotesize(0.987)}
\textbf{Q99996}, P67870, \textbf{P35241}, O60885, P26358, \textbf{Q9UHD2}, Q12923, Q86YT6, \textbf{Q04726}, P61106
\\
\newmidrule
M &
% \textbf{O75439\footnotesize(0.985)},
% P49593{\footnotesize(0.504)},
% P33993{\footnotesize(0.464)},
% P78527{\footnotesize(0.439)},
% \textbf{Q9Y312\footnotesize(0.801)}
P26358, Q9NR30, \textbf{O75439}, Q15056, P61962, P49593, P33993, O60885, \textbf{Q9Y312}, P78527
\\
\newmidrule
NSP7 &
% P62834{\footnotesize(0.685)},
% \textbf{P51148\footnotesize(0.879)},
% P62070{\footnotesize(0.418)},
% \textbf{Q8WTV0\footnotesize(0.854)},
% P53618{\footnotesize(0.350)}
P62834, \textbf{P51148}, P62070, P67870, O14578, \textbf{Q8WTV0}, P53618, Q9BS26, O94973, Q7Z7A1
\\
\newbottomrule
\end{tabular}
%}
\label{tab:covid-link}
\vspace{-8pt}
\end{table}

Besides providing PPI prediction, the proposed model can help by identifying high-confidence candidates for potential human protein targets; this is considered as a link prediction task. When a viral protein (such as \covid M protein) is given as the query, along with a specific relation (such as ``binding'' under the experiment system type of ``Affinity Capture-MS''), \modelname can output a list of most likely protein targets by enumerating the triples with top $f_r(h,t)$ scores. 
The predictions are listed in Table \ref{tab:covid-link}.
It is our observation, \modelname can successfully predict the high-confidence human protein targets in the test set from by \cite{gordon2020sars} among its top predictions (marked as boldfaced entities). %which proves that \modelname can successfully utilize the known facts of \covid targets. 
Other than the proteins in the test set, \modelname can also provide a list of reasonable candidates that possess a relatively high MIST score.
For example, P62834 is one of the top-ranked protein targets of \covid NSP7 by our \modelname, which has a MIST score of 0.658. Diving deep into the facts for P62834, though P62834 is not considered as a high-confidence target by \cite{gordon2020sars}, we observe that both P62834  (RAB1A\_HUMAN) and \covid NSP7 interacts with protein P62820 (RAB1A\_HUMAN). Besides, they are both annotated with the cellular component  GO:0016020 (membrane) and enables molecular function GO:0000166 (nucleotide binding), which are possibly the reasons for \modelname making such prediction with a high rank. 
Furthermore, \modelname's predictions include proteins that are not covered by %the existing experiments in 
\cite{gordon2020sars}, which inspires further scientific research to verify. 
%\todo{add benefit if space allows.}

\begin{figure}[!ht]
    \centering
    \includegraphics[width=0.85\columnwidth]{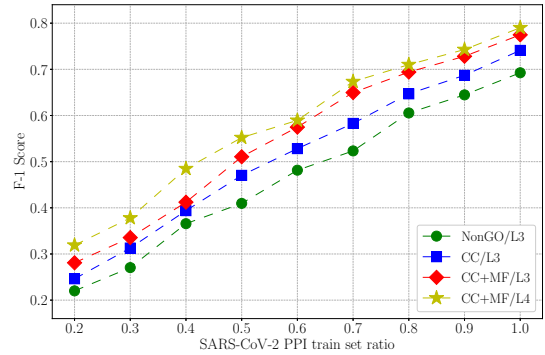}
    \vspace{-10pt}
    \caption{ \modelname performance on different train-set ratios of \covid-Human PPIs.}
    \label{fig:covid-ppi-percentage}
\end{figure}

We further investigate how the information sufficiency of \covid related PPIs in training set affect the performance. We define the train-set ratio parameter as means the proportions of the \covid-Human PPIs that are used for training \modelname and follow the aforementioned evaluation protocol on ``NonGO/S3'', ``CC/S3'', ``CC+MF/S3'' and ``CC+MF/S4'' as input other than the control of \covid-Human PPIs part. We plot the PPI results in Figure \ref{fig:covid-ppi-percentage}. As expected, when the proportion of  \covid-Human PPIs used for training increases from 20\% to 80\%, the F1 score improves from 0.2-0.3 to around 0.8, which strongly confirms that the known \covid-Human PPIs serve as one significant factor to the PPI prediction. Moreover, the more knowledge we know about existing \covid interaction, the more powerful the model is to predict \covid. %However, we also need to point out that the evaluation is still limited by the general availability for \covid resources.
We also observe that the performance is not saturated when the training ratio is approaching  100\%, which possibly results from the fact that as a novel coronavirus, the current known interactions are still very limited. This encourages the scientific communities to unearth more knowledge on \covid; moreover, \modelname has the  potential of bringing about significant advances based on new discoveries.  

%\todo{Visualization of virus protein and part of affected human proteins to see the protein complex in paper \cite{gordon2020sars}.}\junheng{Not easy to observe using DistMult}

%\input{sections/04_Discussion.tex}
\section{Related Work}

% \chelsea{
% 1. Method of representation learning in biological knowledge base: Onto2Vec, and others? \\
% 2. However, Onto2Vec does not provide the capability of joint representation. \\
% 3. describe joint representation learning \\
% 4. mention briefly of the existing methods of joint representation learning \\
% 5. emphasize that they cannot be used or adapted for biological knoweldge base
% }

% following Chelsea's guidelines on related work
% Part 1: Existing representation learning and Onto2Vec
In the past decade, much attention has been paid to representation learning of KBs. 
Methods along this line of research typically encode entities into low dimensional embedding spaces, where the relational inference~\citep{wang2014knowledge}, proximity measures and alignment~\citep{chen2016multilingual} of those entities can be supported in the form of vector algebras. 
Therefore, they provide efficient and versatile methods to incorporate the symbolic knowledge of KGs into statistical learning and inference. 
Some existing approaches focus specifically on computational biology studies~\cite{alshahrani2017neuro, smaili2018onto2vec, chen2019multifaceted, you2015predicting, huang2015using}, which similarly embed features of biological entities within low-dimensional representations. 
One representative work related to ours is Onto2Vec~\cite{smaili2018onto2vec}, in which protein representations are learned by incorporating the full semantic content of gene ontology in the feature learning using Word2Vec \cite{mikolov2013distributed}. 
% Part 2: Refute Onto2Vec
However, Onto2Vec replies on the ontology information, while falls short of capturing the multi-relational semantic facts that are important to characterize the proximity of biological entities. For example, regarding the protein and GO terms, the PPI knowledge and the non-hierarchical relationships between gene ontology entities (such as ``regulates'') are not considered.

% Part 3: joint representation and existing methods
Another thread of related work is joint representation learning for multiple KGs, where embedding models are learned to bridge multiple relational structures for tasks such as entity alignment and type inference.
MTransE \citep{chen2016multilingual} jointly learns a transformation across two separate translational embedding spaces based on one-to-one seed alignment of entities.
Later extensions of this model family, such as KDCoE \citep{chen2018co}, MultiKE \cite{zhang2019multi} and JAPE \citep{sun2017cross}, require additional information of literal descriptions \citep{chen2018co} and numerical attributes of entities \citep{sun2017cross,Trsedya2019attr,zhang2019multi} that are generally not available for biological KB. 
Our recent development on this line of research, i.e. JOIE \citep{hao2019joie} learns a many-to-one mapping between entity embeddings and ontological concept embeddings, and aims at resolving the entity type inference task using the latent space of the type ontology. 
One of the caveats is that JOIE does not specifically incorporate the specificity of concepts in the ontology in the transfer process, for which we find to be particularly beneficial in this problem setting.
% Part 4: Point out existing joint methods are not for bio
Besides, the aforementioned methods are mostly for general encyclopedia KBs (such as Wikidata, DBpedia) and have not been adapted for the purpose the modeling biological KBs.
More specifically, in contrast to these methods, our method features the characterization of more complicated many-to-many associations between proteins and GO terms.
Besides, instead of predicting the alignment of entities, we focus on transferring relational knowledge from one domain to enhance the prediction on the other.

% tentative order
\section{Conclusion}

In this paper, we present a novel model \modelname, that enables end-to-end representation learning for cross-domain biological knowledge bases.
Our approach utilizes the knowledge model to capture structural and relational facts within each domain and motivates the knowledge transfer by alignments among domains.
Extensive experiments on the tasks of PPI type prediction and clustering demonstrate that 
\modelname can successfully leverage complementary knowledge from one domain to another and therefore enable learning entity representation in multiple interrelated and transferable domains in biology.
More importantly, \modelname also provides interaction type predictions on \covid with human protein targets, which potentially brings reliable computational methods seeking new directions on drug design and disease mitigation.

In our main  directions of future research, we plan to enhance  and extend  entity representations by systematically incorporating important multimodal features and annotations. 
For example, primary sequence information and secondary geometric folding features can be modeled simultaneously in protein networks and their combined representation can lead to a  comprehensive understanding that  will greatly benefit many downstream applications.

\begin{acks}
The authors would like to thank the anonymous reviewers for their supportive, insightful and constructive comments.
This work was partially supported by NSF DBI-1565137, NSF III-1705169, NSF CAREER Award 1741634, NSF DGE-1829071, NSF \#1937599, DARPA HR00112090027, NIH R35-HL135772, Okawa Foundation Grant, Amazon Research Award, NEC Research Gift, and Verizon Media Faculty Research and Engagement Program.
\end{acks}

\bibliographystyle{ACM-Reference-Format}
\bibliography{ref.bib}

\end{document}